\documentclass[10pt,a4paper]{article}

\usepackage{graphicx}
\usepackage[latin1]{inputenc}

\title{
\textbf{
Study of the glass transition in the amorphous interlamellar phase of highly crystallized poly(ethylene terephthalate)
}}

\author{
J. Sellarès,
J.A. Diego and
J. Belana\\
\em\small{Departament de Física i Enginyeria Nuclear, Universitat Politècnica de Catalunya,}\\
\em\small{Campus de Terrassa, c. Colom 11, E-08222 Terrassa, Spain.}\\[0.25cm]
}

\date{}

\begin{document}

\maketitle

\begin{abstract}

Poly(ethylene terephthalate) (PET) is a semi--crystalline polymer that can be crystallized up to different degrees heating from the amorphous state. Even when primary crystallization has been completed, secondary crystallization can take place with further annealing and modify the characteristics of the amorphous interlamellar phase. In this work we study the glass transition of highly crystallized PET and in which way it is modified by secondary crystallization. Amorphous PET samples were annealed for 4 hours at temperatures between $140$~$^\circ$C and $180$~$^\circ$C. The secondary crystallization process was monitored by differential scanning calorimetry and the glass transition of the remaining interlamelar amorphous phase was studied by Thermally Stimulated Depolarization Currents measurements. Non--isothermal window polarization is employed to resolve the relaxation in modes with a well--defined relaxation time that are subsequently adjusted to several standard models. Analysis of experimental results reveal that cooperativity is so diminished in crystallized samples with respect to the amorphous material that it can be neglected in the modellization of data. The evolution of the modes during secondary crystallization, once primary crystallization has been completed, gives more weight to lower energy modes. As a consequence, secondary crystallization tends to lower the glass transition temperature of the amorphous interlamellar phase, although remaining noticeably higher than in amorphous samples. Evolution of calorimetric scans of the glass transition are simulated from the obtained results and show the same behavior. It can be concluded that primary and secondary crystallization act in opposite directions on the glass transition temperature of the material even though the effect of secondary crystallization is much slighter. The interpretation of these results in terms of current views about secondary crystallization is discussed. 

\end{abstract}

{\small \noindent Keywords: PET, TSDC, $\alpha$ relaxation, glass transition, amorphous interlamellar phase, secondary crystallization.}

\newpage

\section{Introduction}
\label{intro}

Poly(ethylene terephthalate) (PET) is a widely used semi--crystalline polymer. Just to mention a few applications, it is used to manufacture films, tapes, mouldings, bottles and engineering components. These applications take advantage of the outstanding chemical resistance, thermal stability and mechanical performance of PET.  These properties depend strongly on the microstructure of the material that, on its turn, is determined by the crystallization conditions. Not surprisingly, the relevance of PET applications has spurred a lot of interest about crystallinity in PET. 

In PET, as in other in semi--crystalline polymers, crystalline regions are usually found inside spherulites. These spherical semi--crystalline structures grow during crystallization and tend to fill all the available space.  Spherulites contain plates of highly ordered polymer chains named crystalline lamella. Individual lamella tend to form stacks (sometimes called fibers) that radiate from the center of the spherulite. These stacks are separated by amorphous regions \cite{structure}. Although a small amount of amorphous material can exist between the individual lamella that form the stacks \cite{structure}, most of the amorphous material in the spherulite is placed between the stacks of lamellae. We will refer to this material as the amorphous interlamellar phase. If the material has a low degree of crystallinity, there is also an interspherulitic amorphous phase. This phase is predated by the spherulites as the crystallinity degree grows.

Besides the usual crystallization process that takes place when the material is cooled down slowly from the melt, PET can be crystallized as well heating from the amorphous state (cold crystallization).  This state can be reached if the material is cooled fast enough from the melt (quenching). In fact, the amorphous state can be considered as a supercooled liquid. As the material is heated, chains gain enough mobility to fold and create lamellae more or less in the same way as in the usual crystallization process \cite{structure}. 

Many studies deal about the kinetics of the crystallization process in polymers \cite{secondary}. It has been found that the process can take place in two stages, the so--called primary and secondary crystallization. Primary crystallization is a three--dimensional growth process and it is commonly accepted that it corresponds to the development of the spherulites. Instead, secondary crystallization takes place if the sample is annealed at high temperatures once primary crystallization has been completed (this is when spherulites fill up the avaliable space), and occurs inside the amorphous interlamellar regions \cite{secondary, isothermal}.

There is some controversy on what is really secondary crystallization \cite{isothermal}. It is known that it corresponds to a one--dimensional growth process \cite{secondary} but its exact nature remains unclear. It has been attributed to different causes such as thickening and rearrangement of existing lamellae, insertion of new lamellae inside the stacks, insertion of new stacks as ramifications of existing stacks, creation of randomly oriented new stacks in the interlamellar amorphous phase or connection between stacks with micella structures or folded chains \cite{structure, secondary, isothermal, verma, lin, insight}. There is certain consensus on the fact that as a result of secondary crystallization a dual population of lamellae is created.

A manifestation of secondary crystallization is a minor low--temperature endothermic peak that can be seen by DSC \cite{insight}. This peak appears, when the annealed sample is measured in a non--isothermal scan, at a temperature some degrees higher than the annealing one. It is caused by the fusion of material crystallized by secondary crystallization during the annealing stage \cite{melts}.

The presence of crystalline regions in semi-crystalline polymers is known as well to affect the glass transition of the material. The glass transition temperature is higher in crystallized samples as a consequence of the isolation of the amorphous phase in multiple portions. We must remark that usually it is more difficult to study the glass transition in such systems, as extensive measurable properties (as enthalpy recovery, volume, etc.) are weaker in this case.

It is commonly accepted that the glass transition in polymers is a distributed process. In such processes different parts of the system respond with different relaxation times. A convenient way to study these transitions assumes that the complete relaxation can be modeled by a continuous distribution of relaxation times (DRT). A DRT is thus a set of elementary modes that can reproduce the behavior of the whole process by the superposition of all of them.

Calorimetry and dilatometry have been throughly used to study the glass transition of polymers \cite{structure, wroclaw}, although some complexity arouses in the interpretation of the results because of the distributed nature of the glass transition. Other techniques, such as WAXS or SAXS, are completely unable to detect the amorphous phase in any way \cite{secondary}.

Dielectric techniques have some advantages that make them a worthy option. Almost any change in the material modifies its dielectric properties, so it is possible to study a wide range of phenomena at least indirectly. In the case of the glass transition of polar polymers, the change in mobility of the main polymer chain will result in a change in the polarizability of the material. Dynamic electrical analysis (DEA) is a well-known method that is based on the interaction of an external alternating field with the electric dipoles present in the sample \cite{koji}. 

Among dielectric techniques, thermally stimulated depolarization currents (TSDC) \cite{tsdc} stands out because of its high resolution and low equivalent frequency \cite{hires}. TSDC can be used to study the $\alpha$ relaxation and how it is affected by crystallization \cite{pen}. This relaxation is the dielectric manifestation of the glass transition \cite{tgdin} and, therefore, the measured current comes from the depolarization of the amorphous phase exclusively \cite{belana}. 

The most interesting feature of TSDC is the ability to resolve a complex relaxation into its elementary components. This is done using a procedure known as relaxation map analysis (RMA) \cite{laca}. If the polarizing field, in a TSDC experiment, is on during a large portion of the cooling ramp (conventional polarization) many relaxation processes can be detected in the same thermogram. Instead, if the polarizing field is on in a narrow thermal window, the activated process will behave approximately in an elementary way and will be well described by a single relaxation time $\tau$ (see experimental section for details on a TSDC measurement). 

Analysis of the $\alpha$ relaxation by TSDC gives thus information about the relaxation time, both its mean value, its distribution and cooperativity. A complete and more detailed explanation of the whole process can be found in a previous work of our group \cite{niw}.

The aim of this work is to study the glass transition in highly crystallized PET and, more specifically, how it is affected by the secondary crystallization on the amorphous interlamellar phase. TSDC has been applied to cold--crystallized PET samples and the $\alpha$ relaxation has been studied. Also, DSC scans have been performed to monitor the secondary crystallization process. This kind of study is complementary of the ones focused on the crystalline phase. For this reason, it can give some clues about some controversial points on the morphology of highly crystallized PET, specially on the mechanism that gives rise to secondary crystallization.

\section{Experimental}

\label{exp}

Experiments were carried out on commercial PET, supplied by Autobar, $300$~$\mu$m thick sheets. As received, the material was almost amorphous with less than $3$\% crystallinity degree. From previous works \cite{cold} it was known that $T_g$ for this polymer is approximately $80$~$^\circ$C and that cold crystallization takes place above $100$~$^\circ$C.

Samples were cut in squares of $2$~cm side. Some samples were reserved for as--received measurements and the other ones were annealed. Each sample was annealed for 4~h at one of the following temperatures: $140$~$^\circ$C, $150$~$^\circ$C, $160$~$^\circ$C, $170$~$^\circ$C, $180$~$^\circ$C. After annealing, samples were quenched to room temperature. These samples were used to perform Differential Scanning Calorimetry (DSC) and Thermally Stimulated Depolarization Currents (TSDC) measurements. 

Calorimetric measurements were made with a Mettler TC11 thermoanalyser equipped with a Mettler--20 Differential Scanning Calorimeter module. The calorimeter has been previously calibrated with metallic standards (indium, lead, zinc). To obtain DSC curves, $20$ to $25$~mg portions of the samples were sealed in aluminium pans. Scans begin at $40$~$^\circ$C and end at $300$~$^\circ$C and are performed at a heating rate of $2.5$~$^\circ$C/min. In some experiments, additional thermal treatment was performed in the calorimeter previously to the scan.

Samples for TSDC measurements were prepared coating $1$ cm diameter Al electrodes on both sides of the sample by vacuum deposition. TSDC measurements  were carried out in a non-commercial experimental setup, controlled by an Eurotherm--2416 temperature programmer. Temperature, during measurements,  were measured to an accuracy of $0.1$~$^\circ$C by a J--thermocouple located inside the electrodes (in direct contact with the sample). A Keithley--6512 electrometer has been employed for the current intensity measurements.

Most TSDC experiments have been performed using the non isothermal windowing polarization (NIW) method. According to this method, the sample is continuously cooled from the initial temperature ($T_i$) to the storage temperature ($T_s$) and the polarizing field is applied during the cooling ramp when the temperature of the sample reaches $T_p$ and switched off $\Delta T = 2$~$^\circ$C below $T_p$. In some experiments, referred as conventional polarization in the text, $\Delta T$ has been enlarged significantly, in order to record the relaxation as a whole in spite of not having a unique relaxation time.

The sample remains at $T_s$ for a short storage time ($t_s$) and then it is heated at a constant rate while the TSDC discharge is recorded. The experiment ends at a final temperature ($T_f$). Usually, this final temperature is taken as the $T_i$ of the following experiment. This implies that for a given sample the $T_f$ of an experiment should be lower than the $T_f$ of the previous one, in order to ensure that no relaxation activated in previous experiments is recorded. Since $T_f$ are chosen so that the whole $\alpha$ relaxation is recorded, experiments proceed from higher to lower polarization temperatures.

In all the experiments, $V_p=800$~V, $T_s=30$~$^\circ$C, $t_s=5$~min and the cooling and heating rate is $2.5$~$^\circ$C/min. $T_i$ and $T_f$ have values between $90$~$^\circ$C and $125$~$^\circ$C, well above $T_g$ but, anyway, much lower than the annealing temperature so several experiments can be performed on the same sample without modifying crystalline morphology.

\section{Results and discussion}
\label{resdis}

\subsection{Calorimetric characterization}

In first place we will consider the DSC scan of the material as--received, presented in figure~\ref{figura1}. The plot presents data from just above $T_g$ until the complete fusion of the sample. The crystallization peak can be seen clearly at about $120$~$^\circ$C. At $125$~$^\circ$C most of the crystallization process has been completed, although the process has a long tail that extends onwards. No other feature can be observed in the scan until the fusion peak. This is a broad peak with a maximum at about $250$~$^\circ$C.

When the annealed material is measured by DSC, see figure~\ref{figura2}, the crystallization peak disappears since this process has already taken place during the preparation of the sample. Instead, a smaller but broader endothermic peak appears at a slightly higher temperature than the annealing one. As we consider samples with a higher annealing temperature, the peak appears at progressively higher temperatures. Although the temperature at which the crystallization process ends in the as received material is unclear, as it can be seen in figure~\ref{figura1}, the new peak is not due to the fraction of the tail that remains above the annealing temperature. Otherwise it should have exothermic sign. 

Figure~\ref{annealing} shows DSC scans, that correspond to this minor endothermic peak, with different previous annealing times at $160$~$^\circ$C. Interestingly, the area of the peak is greater for longer annealing times, although most of the process takes place in the first half hour. It can be safely concluded that it is due to the fusion of some structure grown during the annealing stage. More specifically, secondary crystallization \cite{secondary} during the annealing stage is the most probable cause of the new endothermic peak \cite{isothermal, melts}.

The glass transition of crystallized samples is scarcely observable by DSC because the interlamellar amorphous phase yields a much weaker signal than the fully amorphous material. For this reason we will use TSDC to study the glass transition of highly crystallized samples, as explained in the next section.

\subsection{Measurement of the glass transition by TSDC}

It would be interesting to study the effect of annealing in the glass transition of the interlamellar amorphous phase. This can be done tracking the $\alpha$ dielectric relaxation, which takes place only in the amorphous phase and is the dielectric manifestation of the glass transition in polar polymers \cite{tgdin}. TSDC is an idoneous technique to study this relaxation because of its high resolution and low equivalent frequency \cite{hires}.

In figure~\ref{figura3} a comparison is shown between a plot of TSDC experiments on an as--received sample (amorphous) and an annealed one. Conventional polarization is used in both experiments. The as--received sample was polarized from $86$~$^\circ$C  to $40$~$^\circ$C while the crystallized sample was polarized from $106$~$^\circ$C  to $40$~$^\circ$C. In both cases the polarization temperature interval was carefully chosen in order to register the whole relaxation.

It can be seen in these measurements that glass transition of the remaining amorphous fraction of annealed material shows up at higher temperatures and is far more symmetrical than in the case of the almost fully amorphous sample. 

\subsection{Relaxation Map Analysis}

Further information relative to the glass transition can be obtained performing a relaxation map analysis \cite{laca}. In fact, one of the main advantages of the TSDC tecnique is that it allows to study on its own the parts of the mechanism (modes) that give rise to the relaxation as a whole \cite{niw}. To perform the RMA, NIW polarization is employed through a broad range of polarization temperatures in order to study modes with well--defined relaxation time in each TSDC experiment. An example of this kind of measurements is shown in figure~\ref{modes}, where 10 modes around the maximum of the relaxation are plotted for the sample annealed 4 hours at 160~$^\circ$C. 

If the relaxation mechanism has first order kinetics, the calculated depolarization current of each mode can be obtained from the equation
\begin{equation}
 J(T) = {P_0 \over \tau(T)} \; \exp \left[ - {1 \over \beta} 
 \int_{T_0}^T {dT \over \tau(T)} \right],
\end{equation}
where $P_0$ is the initial polarization of the sample, $\tau(T)$ is the relaxation time of the process, $T_0$ is the initial temperature of the experiment and $\beta$ is the heating rate. To modelize the TSDC spectrum $J(T)$, the relaxation time of the $\alpha$ process has been evaluated according to several phenomenological models, that are usually applied to calorimetric and dilatometric measurements. The various kinetic parameters involved were evaluated fitting $J(T)$ by computational methods to the experimental data. A complete and more detailed explanation of the whole process can be found in a previous work \cite{niw}.

TSDC measurements of annealed samples has been fitted to different models (Arrhenius,  Vogel--Tammann--Fulcher and Tool--Narayanaswamy--Moynihan). In figure~\ref{figura4} a comparison of the obtained results is presented. 

The best results are obtained with the Arrhenius model, as the high symmetry of the peaks suggest. In this model the relaxation time $\tau$ is assumed to obey the equation
\begin{equation}
 \tau(T)= \tau_0 \; \exp \left( {E_a \over R T} \right),
\end{equation}
where $E_a$ is the activation energy of the process and $\tau_0$ the pre-exponential factor. It must be noted that this model assumes isolated dipoles and therefore does not take cooperativity into account.

The Vogel--Tammann--Fulcher model, given by the equation 
\begin{equation} 
\tau(T)=\tau_0 \; \exp \left[ E_w \over {R(T-T_{\infty})} \right],
\end{equation}
is not able to improve significantly the results of the Arrhenius model. In fact both fittings are very close, with Arrhenius performing better at low temperatures. Both models are equivalent for $T_{\infty}=0$. In our case, the obtained value for $T_\infty$ is about $110$~K lower than $T_g$, which is a sign that cooperativity can be disregarded. Although there is an overall marginal improvement, it does not justify the inclusion of an additional parameter.

Also, the Tool--Narayanaswamy--Moynihan model \cite{tnm} was employed to modelize the TSDC spectrum, as in \cite{niw}, but the non--linearity parameter was  found to be $1$. As in previous cases, this value represents a non-cooperative relaxation. For this value of the parameter the model is equivalent to the Arrhenius one and identical results are obtained. For this reason it has not been plotted in figure \ref{figura4}.

Aside from the activation energy and the pre--exponential factor, the relative importance (weight) of each mode $N$ can be obtained from the total area of the peak, this is $P_0$, through \cite{laca} 
\begin{equation}
N = P_0 T_p,
\end{equation}
where $P_0$ is multiplied by $T_p$ to take into account the intrinsic dependence of static polarizability on temperature. Nevertheless, at the polarization temperature range employed this correction is very slight.

While it is clear that the effect of primary crystallization is an increase in the value of $T_g$, the effects of secondary crystallization on $T_g$ remain largely unknown. To study this issue we have performed relaxation map analysis on samples annealed at five temperatures at which it has been checked, in the previous section, that secondary crystallization takes place.

Results for each mode of the $\alpha$ relaxation are presented in tables~\ref{table1} to~\ref{table5}. Each table corresponds to a sample previously annealed 4 hours at different temperatures. Results of $P$ in front of $E_a$ for each annealing temperature are also plotted in figure~\ref{figura5}. This figure reveals that the $\alpha$ relaxation is composed by the superposition of modes with a wide range of activation energies centered at $3$~eV.
 
We can discuss the results only in terms of the activation energy because the compensation law \cite{laca} is fulfilled, as it can be seen in figure~\ref{figura6}. This law states a linear relationship between the activation energy and the logarithm of the pre-exponential factor, so it is exactly the same to discuss our results in terms of either quantity.

A significant plot, $N$ in front of $T_p$, can be seen in figure~\ref{ntp}. This figure allows a comparison between data that corresponds to samples previously annealed at different temperatures. High energy modes, that are excited at higher polarization temperatures, tend to loose importance as samples are annealed at higher temperatures while low energy modes, that correspond to lower polarization temperatures, gain relative weight. 

It is interesting to plot the polarization temperature at which a mode is activated in front of the activation energy of the mode. This plot can be seen in figure~\ref{figura7}. From this plot it can be inferred that the modes that reproduce this relaxation have activation energies comprised between $1.25$~eV and $3.75$~eV approximately. It is clear that no modes exist with activation energies outside of this range.

\subsection{Comparison with the glass transition in fully amorphous PET}

These results can be compared with the ones obtained from amorphous PET in a previous work \cite{niw} whose RMA is reproduced in figure~\ref{amorphous}. The most striking difference between the amorphous and crystallized samples lies in the overall shape of the discharge peak, that is much less symmetric in the case of the amorphous material. This lack of symmetry lead to unsatisfactory results if the Arrhenius equation is used to modelize the discharges, opposite to the results obtained with annealed samples. For amorphous samples the VTF and the TNM models provide a significantly closer fit to experimental data \cite{niw}. As a consequence, it can be inferred that cooperativity is much lower in crystallized samples than in amorphous ones.

Another significant difference between amorphous and annealed samples results arose in  the calculated activation energies. In the case of annealed samples this parameter is comprised between $1.25$~eV and $3.75$~eV approximately, as stated above, and a clear upper limit can be inferred from figure~\ref{figura7}. In the case of  amorphous samples no upper limit seems to be present with significant contribution from modes with activation energy of $8$~eV and even higher \cite{niw}. Although obtained values may not be directly comparable because different models have been used to fit each curves, it is significant the presence of this cutoff in annealed samples. This behavior may be a consequence of the limited distance range at which interactions between molecules can propagate during the glass transition, due to the fractioning of the interlamellar amorphous phase. 

This interpretation is  coherent as well with the observed  evolution of the modes due to secondary crystallization. In this case modes with the higher activation energies do not disappear, but low energy modes gain relative weight. Taking into account all these observations, it seems clear that during secondary crystallization some changes occur in the amorphous interlamellar phase that lower slightly the glass transition temperature, unlike primary crystallization that raises clearly the glass transition temperature. This changes will be discussed in section \ref{discuss}.

\subsection{Modelization of DSC curves}

The analysis of TSDC data allows us to predict the shape that would have a calorimetric curve obtained from the glass transition of the interlamellar amorphous phase. This can be done through the fictive temperature, that represents, for a non--equilibrium system, the temperature at which the same system at equilibrium would have the same structural conformation.

To calculate the evolution of the fictive temperature, we assume that the structural relaxation is a first--order distributed process.  This means that the fictive temperature of each mode evolves according to the relaxation time determined in the relaxation map analysis, as
\begin{equation}
\frac{dT_{fi}}{dt} = \frac{T-T_{fi}}{\tau_i},
\label{tfi}
\end{equation}
where $\tau_i$ is the relaxation time and $T_{fi}$ is the fictive temperature, of mode $i$. Since the thermal history is known and in part goes over $T_g$, we can solve easily equation \ref{tfi} for each mode, taking into account that $T_{fi}=T$ when $T>T_g$. In our calculation, $\tau_i$ is obtained from the Arrhenius model and the parameters presented in tables \ref{table1} to \ref{table5}. 

The fictive temperature of the system as a whole, $T_f$, is calculated as the weighted mean of the fictive temperature of all the modes, using $N_i$ as the weight of each mode \cite{niw}.

Once $T_f(t)$ has been calculated for a given thermal history that corresponds to the DSC scan that is being simulated, the normalized calorific capacity $C_p^n$ can be obtained with \cite{cpcalc}
\begin{equation}
C_p^n = \frac{dT_f}{dT}.
\end{equation}

Calculated DSC scans at 2.5~K/min heating rate are presented in figure~\ref{figura8} for differently annealed samples. It can be seen that $T_g$ slightly shifts toward lower temperatures with respect to crystallized samples annealed at lower temperatures, but anyway remaining higher than in amorphous samples. 

This can also be drawn in a more direct way looking for the temperature at which the $\alpha$ relaxation yields a larger current \cite{tgdin}, as seen in figure~\ref{conventional}. It can be seen that in samples annealed at higher temperatures the maximum current, and therefore the glass transition, takes place at slightly lower temperatures. The area of the peaks is also lower for samples annealed at higher temperature. This can be due to a decrease of the amorphous fraction, a decrease of its polarizability or to a combination of both things.

\subsection{Consequences of second crystallization in the glass transition}
\label{discuss}

Up to this point we have presented the effects of secondary crystallization on the glass transition of the amorphous interlamellar phase. However we can take this discussion one step further and analyse the capability of existing models on secondary crystallization to explain these effects. 

It can be inferred, as it is generally assumed, that secondary crystallization takes place in the amorphous interlamellar phase (between the stacks of lamellas). Our results track the glass transition of the amorphous fraction present in the material, that is mainly located between the existing stacks (referred as interlamellar amorphous phase in this work). The fact that secondary crystallization affects so strongly the glass transition of these amorphous fraction indicates  that  it takes place basically in these regions. The formation of new lamellas, from the very scarce amorphous fraction present between the lamellas inside the existing stacks \cite{structure}, does not seem to be, thus, the main contribution to secondary crystallization. 

Other models have been suggested to explain secondary crystallization in these polymers; thickening or rearrangement of existing lamellas or creation of new stacks as ramifications of existing stacks \cite{lin} and creation of randomly oriented new stacks in the interlamellar amorphous phase \cite{structure, isothermal} or connection between stacks with micella structures or folded chains \cite{insight}. According to the presented evolution of the glass transition of annealed samples in this work, all of them are compatible with the observed decrease in the $\alpha$ peak with annealing. However, probably the creation of randomly oriented new stacks in the interlamellar amorphous phase  can better explain  the observed decrease of the relative weight of high activation energy modes when annealing takes place at higher temperatures. In any case we think that only if these new crystalline structures are small enough to be considered randomly and homogeneously distributed, they can effectively affect the relaxation time distribution as a whole.

\section{Conclusions}

The minor low--temperature endothermic peak that appears in annealed samples before the main fusion peak should be attributed to fusion of the fraction of the material that has been crystallized during annealing by secondary crystallization. This crystallized fraction is prone to melt at a slightly greater temperature than the one at which has been crystallized.

On the other hand, TSDC experiments do not reach the annealing temperature in any case so the effect of secondary crystallization in the glass transition can be analyzed by means of TSDC data. The results obtained in TSDC experiments show changes in the $\alpha$ relaxation of the amorphous fraction. 

TSDC results have been fitted to the Arrhenius model since either the Vogel--Tammann--Fulcher and Tool--Narayanaswamy--Moynihan falled back to Arrhenius--like results (in both models Arrhenius is included as a particular case) due to the high symmetry of the TSDC peak.

Results from curve fitting show modes with a limited interval of activation energies in all the cases. Values obtained range from $1.25$~eV for low temperature modes, to $3.75$~eV for the modes that appear at higher temperatures. 

Samples crystallized at higher temperatures show as well a decrease in the fraction of high activation energies modes (those that respond at higher temperatures) in comparison with those modes of lower activation energies.  The overall weight of the modes is thus shifted towards lower activation energies

Secondary crystallization would explain this behavior since it would increase movement restrictions in the amorphous phase, and replace large chain segment movements by smaller ones. As a consequence, the activation energy range would be reduced on the high energy side, as it has been shown in our results.

TSDC experiments also show a displacement of $T_g$ towards lower temperatures, both directly, because the dynamic value of $T_g$ can be associated to the maxima of TSDC plots, and indirectly, reproducing calorimetric data from the obtained dielectric parameters. 

As usual in relaxation map analysis, the compensation law is fulfilled, simplifying the discussion in some degree.

Although secondary crystallization has been investigated by many means, its influence on the dielectric properties of the amorphous interlamellar phase allows us to favor some of the descriptions that have been given to the phenomena, especially those related with the growth of structures in the interlamellar amorphous phase.

\newpage

\pagestyle{empty}

\begin{table}[h]

\caption{Fits for samples annealed $4$~h at $140$~$^\circ$C. \label{table1}}
\begin{center}
\begin{tabular}{|ccccc|}\hline
$T_p$~(K) & $\tau_0$~(s) & $E_a$~(eV) & $P_0$~(C) & $N/N_{max}$ \\ \hline \hline
$369.15$ & $8.26 \times 10^{-51}$ & $3.82$ & $2.16 \times 10^{-09}$ & $0.681$ \\ 
$367.15$ & $4.64 \times 10^{-51}$ & $3.82$ & $2.43 \times 10^{-09}$ & $0.765$ \\ 
$365.15$ & $6.95 \times 10^{-50}$ & $3.72$ & $2.71 \times 10^{-09}$ & $0.845$ \\ 
$363.15$ & $1.14 \times 10^{-47}$ & $3.54$ & $2.97 \times 10^{-09}$ & $0.923$ \\ 
$361.15$ & $9.95 \times 10^{-46}$ & $3.39$ & $3.15 \times 10^{-09}$ & $0.972$ \\ 
$359.15$ & $1.33 \times 10^{-43}$ & $3.22$ & $3.22 \times 10^{-09}$ & $0.989$ \\ 
$357.15$ & $7.35 \times 10^{-40}$ & $2.94$ & $3.27 \times 10^{-09}$ & $1.00$ \\ 
$355.15$ & $6.89 \times 10^{-37}$ & $2.72$ & $3.13 \times 10^{-09}$ & $0.951$ \\ 
$353.15$ & $6.19 \times 10^{-33}$ & $2.43$ & $2.91 \times 10^{-09}$ & $0.879$ \\ 
$351.15$ & $8.05 \times 10^{-30}$ & $2.20$ & $2.55 \times 10^{-09}$ & $0.765$ \\ 
$349.15$ & $2.41 \times 10^{-27}$ & $2.02$ & $2.20 \times 10^{-09}$ & $0.656$ \\ 
$347.15$ & $1.43 \times 10^{-23}$ & $1.75$ & $1.84 \times 10^{-09}$ & $0.546$ \\ 
$345.15$ & $3.73 \times 10^{-22}$ & $1.65$ & $1.48 \times 10^{-09}$ & $0.437$ \\ 
$343.15$ & $4.43 \times 10^{-20}$ & $1.50$ & $1.20 \times 10^{-09}$ & $0.351$ \\ 
$341.15$ & $4.12 \times 10^{-19}$ & $1.42$ & $9.34 \times 10^{-10}$ & $0.273$ \\ 
$339.15$ & $4.70 \times 10^{-18}$ & $1.35$ & $7.24 \times 10^{-10}$ & $0.210$ \\ 
$337.15$ & $2.12 \times 10^{-19}$ & $1.42$ & $5.34 \times 10^{-10}$ & $0.154$ \\ 
$335.15$ & $2.48 \times 10^{-19}$ & $1.41$ & $3.99 \times 10^{-10}$ & $0.114$ \\ 
$333.15$ & $4.62 \times 10^{-18}$ & $1.32$ & $3.51 \times 10^{-10}$ & $0.100$ \\ 
$331.15$ & $7.06 \times 10^{-20}$ & $1.43$ & $2.49 \times 10^{-10}$ & $0.070$ \\ 
\hline
\end{tabular}
\end{center}

\end{table}

\newpage

\begin{table}[h]

\caption{Fits for samples annealed $4$~h at $150$~$^\circ$C. \label{table2}}
\begin{center}
\begin{tabular}{|ccccc|}\hline
$T_p$~(K) & $\tau_0$~(s) & $E_a$~(eV) & $P_0$~(C) & $N/N_{max}$ \\ \hline \hline
$371.15$ & $6.95 \times 10^{-47}$ & $3.55$ & $1.82 \times 10^{-09}$ & $0.608$ \\ 
$369.15$ & $1.22 \times 10^{-47}$ & $3.59$ & $2.00 \times 10^{-09}$ & $0.663$ \\ 
$367.15$ & $2.18 \times 10^{-47}$ & $3.55$ & $2.22 \times 10^{-09}$ & $0.732$ \\ 
$365.15$ & $3.01 \times 10^{-47}$ & $3.53$ & $2.47 \times 10^{-09}$ & $0.809$ \\ 
$363.15$ & $6.04 \times 10^{-49}$ & $3.63$ & $2.66 \times 10^{-09}$ & $0.865$ \\ 
$361.15$ & $5.63 \times 10^{-47}$ & $3.47$ & $2.93 \times 10^{-09}$ & $0.949$ \\ 
$359.15$ & $5.28 \times 10^{-45}$ & $3.32$ & $3.03 \times 10^{-09}$ & $0.976$ \\ 
$357.15$ & $1.31 \times 10^{-42}$ & $3.13$ & $3.10 \times 10^{-09}$ & $0.992$ \\ 
$355.15$ & $3.37 \times 10^{-38}$ & $2.81$ & $3.14 \times 10^{-09}$ & $1.00$ \\ 
$353.15$ & $6.60 \times 10^{-35}$ & $2.57$ & $2.99 \times 10^{-09}$ & $0.949$ \\ 
$351.15$ & $5.33 \times 10^{-30}$ & $2.21$ & $2.74 \times 10^{-09}$ & $0.863$ \\ 
$349.15$ & $1.17 \times 10^{-26}$ & $1.97$ & $2.46 \times 10^{-09}$ & $0.769$ \\ 
$347.15$ & $4.94 \times 10^{-24}$ & $1.78$ & $2.12 \times 10^{-09}$ & $0.660$ \\ 
$345.15$ & $8.50 \times 10^{-21}$ & $1.55$ & $1.84 \times 10^{-09}$ & $0.569$ \\ 
$343.15$ & $1.17 \times 10^{-18}$ & $1.40$ & $1.54 \times 10^{-09}$ & $0.473$ \\ 
$341.15$ & $7.65 \times 10^{-17}$ & $1.27$ & $1.27 \times 10^{-09}$ & $0.390$ \\ 
$339.15$ & $7.22 \times 10^{-15}$ & $1.13$ & $1.10 \times 10^{-09}$ & $0.336$ \\ 
$337.15$ & $3.95 \times 10^{-14}$ & $1.08$ & $9.08 \times 10^{-10}$ & $0.275$ \\ 
$335.15$ & $2.96 \times 10^{-13}$ & $1.01$ & $7.63 \times 10^{-10}$ & $0.229$ \\ 
$333.15$ & $1.49 \times 10^{-13}$ & $1.02$ & $5.97 \times 10^{-10}$ & $0.178$ \\ 
$331.15$ & $3.95 \times 10^{-14}$ & $1.05$ & $4.65 \times 10^{-10}$ & $0.138$ \\ 
$329.15$ & $2.64 \times 10^{-13}$ & $0.99$ & $4.19 \times 10^{-10}$ & $0.124$ \\ 
\hline
\end{tabular}
\end{center}

\end{table}

\newpage

\begin{table}[h]

\caption{Fits for samples annealed $4$~h at $160$~$^\circ$C. \label{table3}}
\begin{center}
\begin{tabular}{|ccccc|}\hline
$T_p$~(K) & $\tau_0$~(s) & $E_a$~(eV) & $P_0$~(C) & $N/N_{max}$ \\ \hline \hline
$369.15$ & $2.80 \times 10^{-49}$ & $3.71$ & $1.87 \times 10^{-09}$ & $0.637$ \\ 
$367.15$ & $3.91 \times 10^{-49}$ & $3.68$ & $2.11 \times 10^{-09}$ & $0.715$ \\ 
$365.15$ & $4.01 \times 10^{-49}$ & $3.66$ & $2.35 \times 10^{-09}$ & $0.792$ \\ 
$363.15$ & $1.94 \times 10^{-49}$ & $3.67$ & $2.57 \times 10^{-09}$ & $0.860$ \\ 
$361.15$ & $1.14 \times 10^{-47}$ & $3.52$ & $2.82 \times 10^{-09}$ & $0.939$ \\ 
$359.15$ & $3.35 \times 10^{-47}$ & $3.47$ & $2.93 \times 10^{-09}$ & $0.969$ \\ 
$357.15$ & $2.34 \times 10^{-44}$ & $3.26$ & $3.02 \times 10^{-09}$ & $0.996$ \\ 
$355.15$ & $2.56 \times 10^{-40}$ & $2.96$ & $3.05 \times 10^{-09}$ & $1.00$ \\ 
$353.15$ & $1.35 \times 10^{-36}$ & $2.68$ & $2.93 \times 10^{-09}$ & $0.954$ \\ 
$351.15$ & $1.17 \times 10^{-32}$ & $2.40$ & $2.66 \times 10^{-09}$ & $0.862$ \\ 
$349.15$ & $2.65 \times 10^{-29}$ & $2.16$ & $2.36 \times 10^{-09}$ & $0.761$ \\ 
$347.15$ & $1.30 \times 10^{-26}$ & $1.96$ & $2.01 \times 10^{-09}$ & $0.642$ \\ 
$345.15$ & $8.66 \times 10^{-24}$ & $1.76$ & $1.67 \times 10^{-09}$ & $0.532$ \\ 
$343.15$ & $1.36 \times 10^{-21}$ & $1.60$ & $1.37 \times 10^{-09}$ & $0.433$ \\ 
$341.15$ & $7.64 \times 10^{-20}$ & $1.48$ & $1.09 \times 10^{-09}$ & $0.342$ \\ 
$339.15$ & $1.07 \times 10^{-19}$ & $1.46$ & $8.52 \times 10^{-10}$ & $0.267$ \\ 
$337.15$ & $9.10 \times 10^{-19}$ & $1.39$ & $6.27 \times 10^{-10}$ & $0.195$ \\ 
$335.15$ & $1.43 \times 10^{-19}$ & $1.43$ & $4.28 \times 10^{-10}$ & $0.132$ \\ 
\hline
\end{tabular}
\end{center}

\end{table}

\newpage

\begin{table}[h]

\caption{Fits for samples annealed $4$~h at $170$~$^\circ$C. \label{table4}}
\begin{center}
\begin{tabular}{|ccccc|}\hline
$T_p$~(K) & $\tau_0$~(s) & $E_a$~(eV) & $P_0$~(C) & $N/N_{max}$ \\ \hline \hline
$367.15$ & $6.42 \times 10^{-49}$ & $3.67$ & $2.02 \times 10^{-09}$ & $0.691$ \\ 
$365.15$ & $1.51 \times 10^{-49}$ & $3.69$ & $2.24 \times 10^{-09}$ & $0.764$ \\ 
$363.15$ & $9.93 \times 10^{-50}$ & $3.69$ & $2.46 \times 10^{-09}$ & $0.835$ \\ 
$361.15$ & $1.53 \times 10^{-48}$ & $3.59$ & $2.71 \times 10^{-09}$ & $0.915$ \\ 
$359.15$ & $9.35 \times 10^{-46}$ & $3.37$ & $2.89 \times 10^{-09}$ & $0.969$ \\ 
$357.15$ & $2.45 \times 10^{-44}$ & $3.25$ & $3.00 \times 10^{-09}$ & $1.00$ \\ 
$355.15$ & $4.91 \times 10^{-41}$ & $3.01$ & $3.01 \times 10^{-09}$ & $0.999$ \\ 
$353.15$ & $9.46 \times 10^{-37}$ & $2.70$ & $2.92 \times 10^{-09}$ & $0.964$ \\ 
$351.15$ & $4.66 \times 10^{-33}$ & $2.43$ & $2.72 \times 10^{-09}$ & $0.892$ \\ 
$349.15$ & $5.17 \times 10^{-29}$ & $2.14$ & $2.44 \times 10^{-09}$ & $0.796$ \\ 
$347.15$ & $6.91 \times 10^{-26}$ & $1.91$ & $2.10 \times 10^{-09}$ & $0.681$ \\ 
$345.15$ & $4.79 \times 10^{-23}$ & $1.71$ & $1.82 \times 10^{-09}$ & $0.586$ \\ 
$343.15$ & $1.62 \times 10^{-20}$ & $1.53$ & $1.51 \times 10^{-09}$ & $0.484$ \\ 
$341.15$ & $6.39 \times 10^{-20}$ & $1.49$ & $1.02 \times 10^{-09}$ & $0.324$ \\ 
$339.15$ & $9.53 \times 10^{-19}$ & $1.40$ & $7.92 \times 10^{-10}$ & $0.251$ \\ 
$337.15$ & $2.73 \times 10^{-18}$ & $1.36$ & $6.04 \times 10^{-10}$ & $0.190$ \\ 
$335.15$ & $1.18 \times 10^{-17}$ & $1.31$ & $4.77 \times 10^{-10}$ & $0.149$ \\ 
\hline
\end{tabular}
\end{center}

\end{table}

\newpage

\begin{table}[h]

\caption{Fits for samples annealed $4$~h at $180$~$^\circ$C. \label{table5}}
\begin{center}
\begin{tabular}{|ccccc|}\hline
$T_p$~(K) & $\tau_0$~(s) & $E_a$~(eV) & $P_0$~(C) & $N/N_{max}$ \\ \hline \hline
$367.15$ & $5.53 \times 10^{-47}$ & $3.52$ & $1.74 \times 10^{-09}$ & $0.618$ \\ 
$365.15$ & $6.01 \times 10^{-49}$ & $3.65$ & $1.90 \times 10^{-09}$ & $0.670$ \\ 
$363.15$ & $5.01 \times 10^{-48}$ & $3.56$ & $2.13 \times 10^{-09}$ & $0.749$ \\ 
$361.15$ & $1.50 \times 10^{-48}$ & $3.58$ & $2.38 \times 10^{-09}$ & $0.830$ \\ 
$359.15$ & $2.61 \times 10^{-48}$ & $3.55$ & $2.57 \times 10^{-09}$ & $0.894$ \\ 
$357.15$ & $3.32 \times 10^{-46}$ & $3.38$ & $2.71 \times 10^{-09}$ & $0.934$ \\ 
$355.15$ & $2.39 \times 10^{-43}$ & $3.17$ & $2.91 \times 10^{-09}$ & $1.00$ \\ 
$353.15$ & $3.05 \times 10^{-40}$ & $2.94$ & $2.82 \times 10^{-09}$ & $0.964$ \\ 
$351.15$ & $8.91 \times 10^{-37}$ & $2.68$ & $2.63 \times 10^{-09}$ & $0.892$ \\ 
$349.15$ & $2.53 \times 10^{-32}$ & $2.36$ & $2.50 \times 10^{-09}$ & $0.844$ \\ 
$347.15$ & $1.22 \times 10^{-28}$ & $2.10$ & $2.14 \times 10^{-09}$ & $0.718$ \\ 
$345.15$ & $3.77 \times 10^{-25}$ & $1.85$ & $1.85 \times 10^{-09}$ & $0.618$ \\ 
$343.15$ & $5.12 \times 10^{-22}$ & $1.63$ & $1.56 \times 10^{-09}$ & $0.519$ \\ 
$341.15$ & $3.95 \times 10^{-20}$ & $1.49$ & $1.27 \times 10^{-09}$ & $0.420$ \\ 
$339.15$ & $1.04 \times 10^{-18}$ & $1.39$ & $1.03 \times 10^{-09}$ & $0.339$ \\ 
$337.15$ & $1.21 \times 10^{-17}$ & $1.31$ & $7.81 \times 10^{-10}$ & $0.255$ \\ 
$335.15$ & $1.31 \times 10^{-17}$ & $1.30$ & $6.21 \times 10^{-10}$ & $0.201$ \\ 
\hline
\end{tabular}
\end{center}

\end{table}

\clearpage

\newtheorem{peudefigura}{Figure}

\begin{peudefigura}
{\rm Power for mass unit of as received material during DSC scan at $2.5$~$^\circ$C/min heating rate.}
\label{figura1}
\end{peudefigura}

\begin{peudefigura}
{\rm Power for mass unit of treated samples during DSC scans at $2.5$~$^\circ$C/min heating rate. Samples were annealed for $4$~h from $140$~$^\circ$C (curve a) to $180$~$^\circ$C (curve e) in $10$~$^\circ$C increment steps.}
\label{figura2}
\end{peudefigura}

\begin{peudefigura}
{\rm Power for mass unit of treated samples during DSC scans at $2.5$~$^\circ$C/min heating rate. Samples were annealed at $160$~$^\circ$C for $15$~min (curve a), $30$~min (curve b), $1$~h (curve c), $2$~h (curve d) and $4$~h (curve e). The scale has been adjusted to show just the minor endothermic peak.} 
\label{annealing}
\end{peudefigura}

\begin{peudefigura}
{\rm Experimental TSDC spectra of the $\alpha$ relaxation of an as received sample (continuous line) and of a sample annealed at $160$~$^\circ$C for $4$~h (dashed line), obtained by conventional polarization.}
\label{figura3}
\end{peudefigura}

\begin{peudefigura}
{\rm Experimental TSDC spectra of the $\alpha$ relaxation obtained by NIW polarization at different $T_p$ from $349.15$~K (curve a) to $371.15$~K (curve l) in $2$~K increment steps. The sample was annealed at $160$~$^\circ$C for $4$~h.}
\label{modes}
\end{peudefigura}

\begin{peudefigura}
{\rm Experimental (circle) and calculated by Arrhenius (dotted line) and WLF (dashed line) TSDC spectrum for $T_p = 359.15$~K. Sample annealed at $160$~$^\circ$C for $4$~h. Parameters employed in the calculation: $\tau_0 = 3.35 \times 10^{-47}$~s and $E_a = 3.47$~eV (Arrhenius), $\tau_0 = 3.76 \times 10^{-14}$~s, $E_w = 0.363$~eV and $T_\infty = 242$~K (WLF). Fits performed between $351.1$~K and $364.1$~K (triangle marks on axis).} 
\label{figura4}
\end{peudefigura}

\begin{peudefigura}
{\rm Obtained distribution of $P_0$ in front of $E_a$. Samples annealed for $4$~h at $140$~$^\circ$C (circle), $150$~$^\circ$C (square), $160$~$^\circ$C (diamond), $170$~$^\circ$C (up triangle) and $180$~$^\circ$C (down triangle).}
\label{figura5}
\end{peudefigura}

\begin{peudefigura}
{\rm Compensation plot: $\ln(\tau)$ in front of $E_a$. Samples annealed for $4$~h at $140$~$^\circ$C (circle), $150$~$^\circ$C (square), $160$~$^\circ$C (diamond), $170$~$^\circ$C (up triangle) and $180$~$^\circ$C (down triangle). The line is just a guide for the eye.}
\label{figura6}
\end{peudefigura}

\begin{peudefigura}
{\rm $N/N_{max}$ for each analyzed mode in front $T_p$. Samples annealed for $4$~h at $140$~$^\circ$C (circle), $150$~$^\circ$C (square), $160$~$^\circ$C (diamond), $170$~$^\circ$C (up triangle) and $180$~$^\circ$C (down triangle).}
\label{ntp}
\end{peudefigura}

\begin{peudefigura}
{\rm Plot of $T_p$ in front of $E_a$ for each analyzed mode. Samples annealed for $4$~h at $140$~$^\circ$C (circle), $150$~$^\circ$C (square), $160$~$^\circ$C (diamond), $170$~$^\circ$C (triangle up) and $180$~$^\circ$C (triangle down).}
\label{figura7}
\end{peudefigura}

\begin{peudefigura}
{\rm Experimental TSDC spectra of the $\alpha$ relaxation obtained by NIW polarization at different $T_p$ from $333.15$~K (curve a) to $353.15$~K (curve k) in $2$~K increment steps \cite{niw}. The sample was almost amorphous (less that 3\% crystallinity degree).}
\label{amorphous}
\end{peudefigura}

\begin{peudefigura}
{\rm Calculated $C_p^n$ of the glass transition assuming a $2.5$~K min$^{-1}$ heating rate and material annealed for $4$~h from $140$~$^\circ$C (curve 1) to $180$~$^\circ$C (curve 3) in $20$~$^\circ$C increment steps.}
\label{figura8}
\end{peudefigura}

\begin{peudefigura}
{\rm Experimental TSDC spectra of the $\alpha$ relaxation of samples annealed for $4$~h at $140$~$^\circ$C (a), $160$~$^\circ$C (b) and $180$~$^\circ$C (c), obtained by conventional polarization.
}
\label{conventional}
\end{peudefigura}

\clearpage

\newcommand{\identificacio}{%

\noindent
J. Sellar\`es, J.A. Diego and J. Belana,\\ ``Study of the glass transition in the amorphous interlamellar phase of highly crystallized poly(ethylene terephthalate)''.

}

\newcommand{\dibuix}[2]{%

\newpage

\pagestyle{empty}

\hbox{}\vspace{2cm}

\begin{center}
\includegraphics[width=10cm]{#1}
\end{center}

\vspace{3cm}

\noindent
{\bf Figure #2}

\identificacio

}

\dibuix{figure01.eps}{\ref{figura1}}

\dibuix{figure02.eps}{\ref{figura2}}

\dibuix{figure03.eps}{\ref{annealing}}

\dibuix{figure04.eps}{\ref{figura3}}

\dibuix{figure05.eps}{\ref{modes}}

\dibuix{figure06.eps}{\ref{figura4}}

\dibuix{figure07.eps}{\ref{figura5}}

\dibuix{figure08.eps}{\ref{figura6}}

\dibuix{figure09.eps}{\ref{ntp}}

\dibuix{figure10.eps}{\ref{figura7}}

\dibuix{figure11.eps}{\ref{amorphous}}

\dibuix{figure12.eps}{\ref{figura8}}

\dibuix{figure13.eps}{\ref{conventional}}


\begin{thebibliography}{10}
\newcommand{\enquote}[1]{``#1''}

\bibitem{structure}
C.~A. Daniels, {\em Polymers: structure and properties\/}, chapter~1, pp.
  1--20, Technomic Publishing, Lancaster (Pennsylvania) (1989).

\bibitem{secondary}
Z.~G. Wang, B.~S. Hsiao, B.~B. Sauer and W.~G. Kampert, {\em Polymer\/} {\bf
  40}, 4615--4627 (1999).

\bibitem{isothermal}
X.~F. Lu and J.~N. Hay, {\em Polymer\/} {\bf 42}, 9423--9431 (2001).

\bibitem{verma}
R.~Verma, H.~Marand and B.~Hsiao, {\em Macromol.\/} {\bf 29}, 7767--7775
  (1996).

\bibitem{lin}
F.~J. Medell\'{\i}n-Rodr\'{\i}guez, P.~J. Phillips and J.~S. Lin, {\em
  Macromol.\/} {\bf 29}, 7491--7501 (1996).

\bibitem{insight}
S.~Tan, A.~Su, W.~Li and E.~Zhou, {\em J. Polym. Sci. Part B: Polym. Phys.\/}
  {\bf 38}, 53--60 (2000).

\bibitem{melts}
S.~Tan, A.~Su, W.~Li and E.~Zhou, {\em Macromol. Rapid Commun.\/} {\bf 19},
  11--14 (1998).

\bibitem{wroclaw}
M.~J. Richardson, {\em Calorimetry and thermal analysis of polymers\/},
  chapter~6, pp. 169--188, Hanser Verlag, Munich Vienna New York (1994).

\bibitem{koji}
K.~Fukao and Y.~Miyamoto, {\em Phys. Rev. Lett.\/} {\bf 79}, 4613--4616 (1997).

\bibitem{tsdc}
R.~Chen and Y.~Kirsh, {\em Analysis of thermally stimulated processes\/},
  chapter~3, pp. 60--81, Pergamon, Oxford, 1st edition (1981).

\bibitem{hires}
G.~Teyss\`edre and C.~Lacabanne, {\em J. Phys. D: Appl. Phys.\/} {\bf 28},
  1478--1487 (1995).

\bibitem{pen}
J.~C. Ca{\~n}adas, J.~A. Diego, J.~Sellar\`es, M.~Mudarra, J.~Belana,
  R.~D\'{\i}az-Calleja and M.~J. Sanchis, {\em Polymer\/} {\bf 41}, 2899--2905
  (2000).

\bibitem{tgdin}
J.~Belana, P.~Colomer, S.~Montserrat and M.~Pujal, {\em J. Macromol. Sci. Part
  B Phys.\/} {\bf 23}, 467--481 (1984).

\bibitem{belana}
J.~Belana and P.~Colomer, {\em J. Mat. Sci.\/} {\bf 26}, 4823--4828 (1991).

\bibitem{laca}
A.~Bernes, G.~Teyss\`edre, S.~Mezghani and C.~Lacabanne, {\em Dielectric
  spectroscopy of polymeric materials\/}, chapter~8, pp. 2227--2258, American
  Chemical Society, Washington (DC) (1997).

\bibitem{niw}
J.~A. Diego, J.~Sellar\`es, A.~Aragoneses, M.~Mudarra, J.~C. Ca{\~n}adas and
  J.~Belana, {\em J. Phys. D: Appl. Phys.\/} {\bf 40}, 1138--1145 (2007).

\bibitem{cold}
J.~C. Ca{\~n}adas, J.~A. Diego, J.~Sellar\`es, M.~Mudarra and J.~Belana, {\em
  Polymer\/} {\bf 41}, 8393--8400 (2000).

\bibitem{tnm}
C.~T. Moynihan, A.~J. Easteal, J.~Wilder and J.~Tucker, {\em J. Phys. Chem.\/}
  {\bf 78}, 2673--2677 (1974).

\bibitem{cpcalc}
I.~M. Hodge and A.~R. Berens, {\em Macromol.\/} {\bf 14}, 1598--1599 (1981).

\end{thebibliography}
\end{document}